\documentclass[cmfonts,a4paper]{aipproc}
\layoutstyle{8d}
\usepackage{amssymb,amsmath,amsthm,amscd,verbatim}
\usepackage[mathcal]{euler}
 \font\palba=pplb at 20pt
 \font\palo=pplr at 14pt
 \font\paloa=pplr at 14pt

\title{\palba Maximum Probability and Maximum Entropy methods: Bayesian interpretation}

\author{\palo M. Grendar, Jr. and \underline{M. Grendar}}
{
 address = {Institute of Mathematics and Computer Science of Mathematical Institute of
 Slovak Academy of Sciences (SAS) and of Matej Bel University, Severn\'a ulica 5, 974 00
        Bansk\'a Bystrica, Slovakia \&
 Institute of Measurement Science of SAS,
           D\'ubravsk\'a cesta 9, 841 04 Bratislava,  Slovakia. umergren@savba.sk
           }}

\def\vc{\boldsymbol}
\def\vcs{\boldsymbol}

\begin{document}
\begin{abstract}
(Jaynes') Method of (Shannon-Kullback's) Relative Entropy
Maximization (REM or  MaxEnt) can be - at least in the discrete case
- according to the Maximum Probability Theorem (MPT)  viewed as an
asymptotic instance of the Maximum Probability method (MaxProb). A
simple bayesian interpretation of MaxProb is given here. MPT carries
the interpretation over into REM.
\end{abstract}

\maketitle

\section{\paloa  Introduction}

Relationship of the Method of (Shannon-Kullback's) Relative Entropy
Maximization (REM or MaxEnt) and Bayesian Method is notoriously
peculiar. The two methods of induction are viewed as unrelated at
all, or opposed, or identical in some circumstances, or one as a
special case of the other one (see \cite{Jaynes}).

As it was noted,  a finding  that REM can be viewed as an asymptotic
instance of Maximum Probability method (MaxProb, cf. \cite{ggwhat})
implies that MaxProb/REM/MaxEnt cannot be in conflict with Bayes'
Theorem (cf. \cite{ggmyst}).

A beautiful, simple (yet in some extent overlooked) bayesian
interpretation of REM which operates on the level of samples and
employs Conditioned Weak Law of Large Numbers (CWLLN) was suggested
and elaborated at \cite{Csiszar}. Csisz\'ar's original argument
together with the Maximum Probability Theorem (MPT, see
\cite{ggwhat}, Thm 1), inspired a bayesian interpretation of MaxProb
and REM methods, which we
intend to present here. 

\section{\paloa Terminology and notation}

Let $\mathcal{X} \triangleq \{x_1, x_2, \dots, x_m\}$ be a discrete
finite set called support, with $m$ elements and let $\{X_l, l = 1,
2, \dots, n\}$ be a sequence of size $n$ of identically and
independently drawn random variables taking values in $\mathcal{X}$.

A type $\vcs\nu_n \triangleq [n_1, n_2, \dots, n_m]/n$ is an
empirical probability mass function which can be based on sequence
$\{X_l, l = 1, 2, \dots, n\}$. Thus, $n_i$ denotes number of
occurrences of $i$-th element of
$\mathcal{X}$ in the sequence. 

Let $\mathcal{P(X)}$ be a set of all probability mass functions
(pmf's) on $\mathcal{X}$. Let $\mathcal\Pi_n \subseteq
\mathcal{P(X)}$ be a set of all types $\vcs\nu_n$, and let $\mathcal
H_n \subseteq \mathcal\Pi_n$.

Let the supposed source of the sequences (and hence also of types)
be $\vc q \in \mathcal{P(X)}$.

Let ${\pi}(\vcs\nu_n)$ denote the probability that  $\vc q$ will
generate type $\vcs\nu_n$, ie. ${\pi}(\vcs\nu_n) \triangleq
\frac{n!}{n_1!\,n_2!\,\dots\, n_m!} \prod_{i =1}^m q_i^{n_i}$.

\section{\paloa Bayesian interpretation of Maximum Probability method}

Bayesian recipe prescribes to update prior distribution
(information) by an evidence via Bayes' Theorem (BT) to get a
posterior distribution. Usually bayesians use BT to update prior
distribution of a parameter by evidence which has form of random
sample and obtain posterior distribution of the parameter, given the
sample. Then it is customary to select the value of parameter at
which the posterior distribution attains its maximum (i.e. mode) and
perform further inference.

The bayesian recipe and \cite{Csiszar} will be followed here on a
different level. A prior distribution of types will be updated via
BT by data of special form. Then the maximum aposteriori type will
be searched out.

The bayesian updating  will be carried out in four steps:

Step 1: Select a probability mass function $\vc q$ which could be
the best guess of source of types $\vcs\nu_n$. It will specify a
prior probability $P(\vcs\nu_n)$ of type by the following simple
scheme: $P(\vcs\nu_n) \equiv \pi(\vcs\nu_n)$. Thus, $\pi(\vcs\nu_n)$
is the apriori distribution of types, which is going to be updated
once an evidence (data) will become
available. 

Step 2: The data arrive in rather special form: they specify a set
$\mathcal{H}_n$ of types $\vcs\nu_n$ (which were observed, or
'feasible' in some general way). In other words, the evidence is
that types which do not belong to $\mathcal{H}_n$ cannot be
observed, or are 'not feasible'.

Step 3: Use Bayes' Theorem to update the prior probability of type
$\pi(type = \vcs\nu_n)$ by the evidence "$type \in \mathcal{H}_n$"
to obtain the posterior probability $P(type = \vcs\nu_n | type \in
\mathcal{H}_n)$ that type is equal to $\vcs\nu_n$ given that it
conforms with the evidence (i.e. belongs to $\mathcal{H}_n$).

\begin{equation*}
\begin{split}
P(type = \vcs\nu_n | type \in \mathcal{H}_n)  = \qquad \\
\frac{P(type \in \mathcal{H}_n| type = \vcs\nu_n) \, \pi(type =
\vcs\nu_n)}{P(type \in \mathcal{H}_n)}\qquad
\end{split}
\end{equation*}

Note that $P(type \in \mathcal{H}_n| type = \vcs\nu_n)$ is $0$ if
$\vcs\nu_n \notin \mathcal{H}_n$ and $1$ otherwise.  Thus, for
$\vcs\nu_n \in \mathcal{H}_n$ the aposteriori probability is

\begin{equation*}
P(type = \vcs\nu_n | type \in \mathcal{H}_n) = \frac{\pi(type =
\vcs\nu_n)}{P(type \in \mathcal{H}_n)}
\end{equation*}

Obviously, $P(type \in \mathcal{H}_n) = \sum_{\vcs\nu^n \in
\mathcal{H}_n} \pi(\vcs\nu^n)$.

Step 4: The type(s) with the highest value of the posterior
probability (MAP type) is to be searched out. Since types which do
not belong to $\mathcal{H}_n$ have zero posterior probability, a
search for the MAP type can be restricted to types which belong to
$\mathcal{H}_n$. So, the MAP type $\hat{\vcs\nu}_n$ is
$$
\hat{\vcs\nu}_n \triangleq \arg \max_{\vcs\nu_n \in \mathcal{H}_n}
P(type = \vcs\nu_n | type \in \mathcal{H}_n)
$$
Since, for fixed $n$ and any $\vcs\nu_n$, $P(type \in
\mathcal{H}_n)$ is a constant, the MAP type turns to be
\begin {equation}
\hat{\vcs\nu}_n = \arg \max_{\vcs\nu_n \in \mathcal{H}_n} \pi(type =
\vcs\nu_n)
\end{equation}

Thus the MAP type $\hat{\vcs\nu}_n$ is just the type in
$\mathcal{H}_n$ which has the highest value of the prior
probability.

Here it stops.

Observe that (1) is identical with prescription of the Maximum
Probability (MaxProb) method (cf. \cite{ggwhat}). Thus the above
reasoning provides its bayesian interpretation.

\section{\paloa How does it relate to REM/MaxEnt?}

Via Maximum Probability Theorem (MPT, see \cite{ggwhat}, Thm 1 and
\cite{ggnonlin}).

Before stating MPT, $I$-projection has to be defined. $I$-projection
$\hat{\vc p}$ of $\vc q$ on  set $\mathcal{\Pi} \subseteq
\mathcal{P(X)}$ is such $\hat{\vc p} \in \mathcal{\Pi}$ that
$I(\hat{\vc p}\| \vc q) = \inf_{\vc p \in \mathcal{\Pi}} I(\vc
p\|\vc q)$, where\footnote{There, $\log 0 = - \infty$, $\log
\frac{b}{0} = + \infty$, $0 \cdot (\pm\infty) = 0$, conventions are
assumed.  Throughout the paper $\log$ denotes the natural
logarithm.} $I(\vc p\| \vc q) \triangleq \sum_{\mathcal X} p_i \log
\frac{p_i}{q_i}$ is the $I$-divergence. $I$-divergence is  known
under various other names: Kullback-Leibler's distance, KL number,
Kullback's directed divergence, etc. When taken with minus sign it
is known as (Shannon-Kullback's) relative entropy.

{\bf (MPT)}\footnote{Originally MPT was stated with unique
$I$-projection case in mind. Its proof however readily allows to
state it in general form (see \cite{ggnonlin}). Since the issue of
uniqueness is at this Section irrelevant the MPT will be stated at
its original form.} {\it  Let {\rm differentiable\/} constraint
$F({\vcs{\nu}_n}) = 0$ define feasible set of types $\mathcal{H}_n$
and let $\mathcal{H} \triangleq \{\vc p: F({\vc{p}}) = 0\}$ be the
corresponding feasible set of probability mass functions. Let
$\hat{\vcs\nu}_n \triangleq \arg\, \max_{\vcs\nu_n \in
\mathcal{H}_n} \pi(\vcs\nu_n)$. Let $\hat{\vc p}$ be $I$-projection
of $\vc q$ on $\mathcal{H}$. And let $n\rightarrow\infty$. Then
$\hat{\vcs\nu}_n = \hat{\mathbf{p}}$. }

MPT shows that REM is an asymptotic instance of MaxProb method. Thus
MPT carries the bayesian interpretation of MaxProb over into
REM/MaxEnt. Hence, $I$-projection is just the MAP type which results
from the bayesian updating which was described at the previous
Section, in the case of sufficiently large $n$.

To sum up: Whenever $n$ is sufficiently large and prior will be
assigned to types $\vcs\nu_n$ as in the Step 1, and new data will
take form as in the Step 2, and the prior will be updated by the
data via BT as in the Step 3, and MAP type will be searched out as
in the Step 4, then the MAP type will be nothing but the REM
$I$-projection of $\vc q$ on $\mathcal{H}$.

\section{\paloa Discussion}

Why MAP? Why not say median
aposteriori type? The MAP type becomes when $n \rightarrow \infty$
just the $I$-projection. If the $I$-projection is unique then
Conditioned Weak Law  of Large Numbers (CWLLN, cf. \cite{Vasicek},
\cite{VC}, \cite{G}, \cite{CsiszarS}, \cite{LPS}, \cite{LS2},
\cite{LN}) can be invoked. If read in the above bayesian manner, it
says
 that any other type/distribution than $I$-projection has asymptotically zero posterior
probability. So, this is why MAP and not median. However, what if
there are multiple $I$-projections? Obviously, the bayesian
interpretation of MaxProb is valid regardless of the number of MAP
types. MPT in its general form (cf. \cite{ggnonlin}) covers also the
case of multiple MaxProb types and claims that they converge to
$I$-projections. Then one can either recall Entropy Concentration
Theorem (cf. \cite{ggnonlin}) or invoke an extension of CWLLN which
covers also the case of multiple $I$-projections (cf. \cite{ggaei})
-- to answer the "Why MAP" question in the general case.

\section{\paloa Concluding note}

Originally (cf. \cite{ggwhat}), MaxProb was presented as a method
which looks in $\mathcal{H}_n$ for a type $\hat{\vcs\nu}_n = \arg
\max_{\vcs\nu_n \in \mathcal{H}_n}$ $\pi(\vcs\nu_n)$  which the
'prior' generator $\vc q$  can generate with the highest
probability. The word 'prior' was used merely to mean that the
generator is selected before the data arrive. Alternatively, since
un\-con\-stra\-ined ma\-xi\-mi\-za\-ti\-on of the conditional
probability $P(type = \vcs\nu_n| type \in \mathcal{H}_n)$ reduces to
maximization of $\pi(\vcs\nu_n)$ constrained to $\vcs\nu_n \in
\mathcal{H}_n$, MaxProb could be interpreted as search for the type
with the highest value of the conditional probability. The third,
bayesian interpretation of MaxProb -- inspired by \cite{Csiszar} --
was given here. Obviously, MPT stands regardless of what is the
preferred interpretation of MaxProb.

\section{\paloa Acknowledgements}

Supported by the grant VEGA 1/0264/03 from the Scientific Grant
Agency of the Slovak Republic. It is a pleasure to thank Viktor
Witkovsk\'y for valuable discussions and comments on earlier version
of this work. The thanks extend to Ale\v s Gottvald and George
Judge. Lapses are mine.

\renewcommand\refname

\bigskip\bigskip

{\small Corrections wrt Version 1: $i)$ Obviously, $P(type \in
\mathcal{H}_n)$ is not given as a ratio of the number of types in
$\mathcal{H}_n$ to the number of all types in $\mathcal{\Pi}_n$;
rather it is $P(type \in \mathcal{H}_n) = \sum_{\vcs\nu^n \in
\mathcal{H}_n} \pi(\vcs\nu^n)$. The rest of argument remains
untouched by the gross lapse. $ii)$ The second question from
Discussion (Sect. 5) from the Version 1 is not included here.

\smallskip

In the form of Version 1, the paper appeared as M. Grendar, Jr. and
M. Grendar, Maximum Probability and Maximum Entropy methods:
Bayesian interpretation, in {\it Bayesian Inference and Maximum
Entropy methods in Science and Engineering}, G. Erickson and Y. Zhai
(eds.),
 AIP, Melville, pp. 490 -495, 2004.


}

\end{document}